\documentstyle[sprocl,epsfig]{article}

\def\be{\begin{equation}}
\def\ee{\end{equation}}
\def\bea{\begin{eqnarray}}
\def\eea{\end{eqnarray}}

\begin{document}

\title{Confinement Physics in Quantum Chromodynamics}

\author{H.~Suganuma, H.~Ichie, K.~Amemiya and A.~Tanaka}
\address{Research Center for Nuclear Physics (RCNP), Osaka University \\ 
Ibaraki, Osaka 567-0047, Japan \\
E-mail: suganuma@rcnp.osaka-u.ac.jp}


\maketitle\abstracts{ 
We study the confinement physics in QCD in the maximally abelian 
(MA) gauge using the SU(2) lattice QCD, based on the 
dual-superconductor picture. In the MA gauge, 
off-diagonal gluon components are forced to be small, 
and the off-diagonal angle variable $\chi_\mu(s)$ tends to
be random.
Within the random-variable approximation for $\chi_\mu(s)$,
we analytically prove the perimeter law of the off-diagonal gluon 
contribution to the Wilson loop in the MA gauge, which leads to
abelian dominance on the string tension.
To clarify the origin of abelian dominance for the long-range 
physics, we study the charged-gluon propagator 
in the MA gauge using the lattice QCD, 
and find that the effective mass $m_{ch} \simeq 0.9 {\rm GeV}$ 
of the charged gluon is induced by the MA gauge fixing.
In the MA gauge, 
there appears the macroscopic network of the monopole world-line 
covering the whole system, which would be identified as 
monopole condensation at a large scale. 
To prove monopole condensation in the field-theoretical manner, 
we derive the inter-monopole potential from the dual Wilson loop 
in the monopole part of QCD, 
which carries the nonperturbative QCD aspects, in the MA gauge. 
The dual gluon mass is evaluated as $m_B \simeq $0.5GeV 
in the monopole part in the infrared region, which is the evidence of 
the dual Higgs mechanism by monopole condensation. 
}

\section{QCD and Dual Superconductor Picture for Confinement}

Quantum chromodynamics (QCD) is the fundamental theory 
of the strong interaction. 
In spite of the simple form of the QCD lagrangian 
\bea
{\cal L}_{\rm QCD}=-{1 \over 2} {\rm tr} G_{\mu\nu}G^{\mu\nu}
+\bar q (i\not D-m_q) q, 
\eea
it miraculously provides quite various phenomena 
like color confinement, chiral symmetry breaking, 
nontrivial topologies, quantum anomalies and so on.
It would be interesting to compare QCD with the history of 
the Universe, because a quite simple `Big Bang' 
also created everything including galaxies, stars and 
living things.
Therefore, QCD can be regarded as an interesting 
miniature of the Universe. 
This is the most attractive point of the QCD physics.
In this paper, we will slightly touch the confinement physics in QCD.

The Regge trajectory of hadrons and the lattice QCD show that 
the confinement force between the color-electric charges is 
characterized by the universal physical quantity of the string tension 
$\sigma \simeq 1{\rm GeV/fm}$ and is brought 
by the {\it one-dimensional squeezing} of the color-electric flux 
in the QCD vacuum. 
Based on the electro-magnetic duality, 
Nambu proposed the {\it dual superconductor picture} for quark 
confinement in 1974.$^1$ 
In this picture, such a squeezing of the color-electric flux 
between quarks is realized by the {\it dual Meissner effect,} 
as the result of {\it color-magnetic monopole condensation,} 
which is the dual version of electric-charge condensation 
in the superconductor.
However, there are {\it two large gaps} between QCD and the 
dual-superconductor picture.$^2$

\ 

\noindent
\begin{minipage}{.5cm}
(1)~\\~\\
\end{minipage}
\begin{minipage}{11.3cm}
This picture is based on the abelian gauge theory subject to the 
Maxwell-type equations, where electro-magnetic duality is manifest, 
while QCD is a nonabelian gauge theory. 
\end{minipage}
\noindent
\begin{minipage}{.5cm}
(2)~\\~\\
\end{minipage}
\begin{minipage}{11.3cm}
The dual-superconductor scenario requires condensation of 
(color-) magnetic monopoles as the key concept, while QCD does not 
have such a monopole as the elementary degrees of freedom.
\end{minipage}

\ 

\noindent
These gaps can be simultaneously fulfilled by 
the use of the {\it 't~Hooft abelian gauge fixing,} 
the partial gauge fixing which only remains 
abelian gauge degrees of freedom in QCD.$^3$ 
The abelian gauge fixing reduces QCD to 
an abelian gauge theory, where the off-diagonal gluon 
behaves as a charged matter field 
similar to $W^\pm_\mu$ in the Standard Model and provides 
a color-electric current in terms of the residual abelian gauge 
symmetry.
As a remarkable fact in the abelian gauge, 
{\it color-magnetic monopoles} appear as {\it topological objects} 
corresponding to the nontrivial homotopy group 
$\Pi_2($SU($N_c$)/U(1)$^{N_c-1})$
$=${\bf Z}$^{N_c-1}_\infty$ 
in a similar manner to the GUT monopole.$^{3-5}$

Here, let us consider the appearance of monopoles in terms of 
the gauge connection.$^{2,6}$ 
In the general system including the singularity such as the 
Dirac string, the field strength is defined as 
\be
G_{\mu\nu}\equiv \frac{1}{ie}
([\hat D_\mu, \hat D_\nu]-[\hat \partial_\mu, \hat \partial_\nu]),
\ee
which takes a form of the difference 
between the covariant derivative operator 
$\hat D_\mu \equiv \hat \partial_\mu+ieA_\mu(x)$ 
and the derivative operator $\hat \partial_\mu$ 
satisfying $[\hat \partial_\mu, f(x)]=\partial_\mu f(x)$.
By the general gauge transformation with the gauge function $\Omega$, 
$\hat D_\mu$ is transformed as 
$\hat{D}_\mu \rightarrow \hat{D}_\mu^\prime \equiv \Omega \hat{D}_\mu \Omega^{\dagger}$,
and $G_{\mu\nu}$ is transformed as 
\bea
G_{\mu\nu} \rightarrow 
{G'}_{\mu\nu} &\equiv& \Omega G_{\mu\nu} \Omega^\dagger=
\frac{1}{ie}([\hat {D'}_\mu, \hat {D'}_\nu]-
\Omega[\hat \partial_\mu, \hat \partial_\nu]
\Omega^\dagger) \cr
&=&
\partial_\mu 
{A'}_\nu-\partial_\nu {A'}_\mu+ie[{A'}_\mu, {A'}_\nu]
+\frac{i}{e}\Omega[\partial_\mu,\partial_\nu]\Omega^\dagger. 
\eea
The last term remains only for the {\it singular gauge transformation}, 
and can provide the Dirac string in the abelian gauge sector.
For a singular ${\rm SU}(N_c)$ gauge function, 
the last term leads to breaking of the abelian Bianchi identity 
and monopoles in the abelian gauge.$^{2,6}$ 

\noindent
Thus, QCD in the abelian gauge is an abelian gauge theory 
including both the electric current $j_\mu$ and 
the magnetic current $k_\mu$, 
and can provide the theoretical basis of the dual-superconductor scheme 
for the confinement mechanism.

\section{Maximally Abelian Gauge, 
Abelian Dominance and Global Network of 
Monopole Current in MA Gauge in Lattice QCD}

In the Euclidean QCD, 
the maximally abelian (MA) gauge is 
defined by minimizing 
\be 
R_{\rm off} [A_\mu ( \cdot )] \equiv \int d^4x {\rm tr}
[\hat D_\mu ,\vec H][\hat D_\mu ,\vec H]^\dagger
={e^2 \over 2} \int d^4x \sum_\alpha  |A_\mu ^\alpha (x)|^2, 
\ee
where $\hat D_\mu \equiv \hat \partial_\mu+ieA_\mu $ denotes 
the ${\rm SU}(N_c)$ covariant derivative and 
the Cartan decomposition $A_\mu (x)=\vec A_\mu (x) \cdot \vec H 
+\sum_\alpha A_\mu^\alpha (x)E^\alpha $ is used. 
{\it In the MA gauge, the off-diagonal gluon component 
is forced to be as small as possible by the gauge transformation, 
and therefore the gluon field $A_\mu (x) \equiv A_\mu ^a(x)T^a$ 
closely resembles the abelian gauge field 
$\vec A_\mu (x) \cdot \vec H$.}$^{2,6}$ 

In the MA gauge, $G \equiv {\rm SU}(N_c)_{\rm local}$ 
is reduced into $H \equiv {\rm U(1)}_{\rm local}^{N_c-1} 
\times {\rm Weyl}_{\rm global}^{N_c}$,
where the {\it global Weyl symmetry} is the subgroup of ${\rm SU}(N_c)$ 
relating the permutation of the $N_c$ bases.$^7$ 
Since the covariant derivative $\hat D_\mu$ obeys the 
adjoint gauge transformation, 
the MA gauge fixing condition is found to be$^2$ 
\be
[\vec H, [\hat D_\mu , [\hat D_\mu , \vec H]]]=0,
\ee
which leads 
the partial gauge fixing on $G/H$. 

In the Euclidean lattice formalism,
the MA gauge is defined by maximizing 
the diagonal element of the link variable 
$U_\mu (s) \equiv \exp\{iaeA_\mu (s)\}$,$^{2,8}$ 
\be
R_{\rm diag} [U_\mu ( \cdot )] \equiv \sum_{s,\mu } 
{\rm tr} \{U_\mu (s) \vec H U_\mu^\dagger (s) \vec H \}.
\ee
The ${\rm SU}(N_c)$ link variable is factorized corresponding to the 
Cartan decomposition $G/H \times H$ as 
$U_\mu (s)=M_\mu (s)u_\mu (s)$ with 
$
M_\mu (s) \equiv \exp\{i\Sigma _\alpha \theta _\mu ^\alpha (s)E^\alpha \}
$ and 
$
u_\mu (s) \equiv \exp\{i \vec \theta _\mu (s) \cdot \vec H \}
$.
Here, the {\it abelian link variable} 
$u_\mu (s) \in H \equiv {\rm U(1)}^{N_c-1}$ behaves as 
the abelian gauge field, and the off-diagonal factor 
$M_\mu (s) \in G/H$ behaves as the charged matter field 
in terms of the residual abelian gauge symmetry 
${\rm U(1)}^{N_c-1}_{\rm local}$.
In the lattice formalism, the abelian projection 
is defined by the replacement as 
$
U_\mu (s) \in G ~ \rightarrow ~ u_\mu (s) \in H. 
$

Abelian dominance and monopole dominance for NP-QCD 
(confinement$^9$, D$\chi $SB$^{10}$, instantons$^{7,11}$) are 
the remarkable facts observed in the lattice QCD 
in the MA gauge.
Here, we summarize the QCD system in the MA gauge 
in terms of abelian dominance, monopole dominance and 
extraction of the relevant mode for NP-QCD.

\noindent
\begin{minipage}{.5cm}
(a)~\\
\end{minipage}
\begin{minipage}{11.3cm}
Without gauge fixing, it is difficult to extract 
relevant degrees of freedom for NP-QCD. 
All the gluon components equally contribute to NP-QCD.
\end{minipage}

\noindent
\begin{minipage}{.5cm}
(b)~\\~\\~\\~\\~\\~\\~\\~\\
\end{minipage}
\begin{minipage}{11.3cm}
In the MA gauge, QCD is reduced into an abelian gauge theory 
including the electric current $j_\mu $ and the magnetic current $k_\mu $.
The diagonal gluon $\vec{A_\mu}\cdot\vec{H}$
behaves as the abelian gauge field, 
and the off-diagonal gluon 
behaves as the charged matter field in terms of 
the residual abelian gauge symmetry. 
In the MA gauge, the lattice QCD shows {\it abelian dominance} 
for NP-QCD:
only the diagonal gluon is relevant for NP-QCD, 
while off-diagonal gluons do not contribute to NP-QCD. 
In the confinement phase of the lattice QCD, 
there appears the {\it global network of the monopole world-line 
covering the whole system in the MA gauge} 
as shown in Fig.1. 
\end{minipage}

\noindent
\begin{minipage}{.5cm}
(c)~\\ ~\\ ~\\ ~\\ ~\\ ~\\ 
\end{minipage}
\begin{minipage}{11.3cm}
The diagonal gluon can be decomposed into the 
``photon part'' and the ``monopole part'', 
corresponding to the separation of $j_\mu$ and $k_\mu$.
The monopole part carries the monopole current $k_\mu $ only, 
{\it i.e.} $j_\mu  \simeq 0$. 
The photon part 
carries the electric current $j_\mu $ only, {\it i.e.} 
$k_\mu  \simeq 0$. 
In the MA gauge, the lattice QCD shows {\it monopole dominance} 
for NP-QCD:
the monopole part leads to NP-QCD, 
while the photon part seems trivial like QED and 
does not contribute to NP-QCD. 
\end{minipage}

\noindent 
Thus, monopoles in the MA gauge can be regarded as 
the relevant collective mode for NP-QCD, 
and {\it formation of the global network of the monopole current 
seems to mean ``monopole condensation'' in the infrared 
scale.}$^{2,12}$

\ 

\begin{figure}[htb]
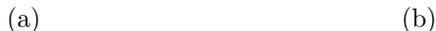

\hspace{1cm}
\hspace{1cm}
$~$\hspace{3cm}(a)\hspace{4.8cm}(b)
          \caption{ 
The monopole world-line projected into ${\bf R}^3$ 
in the MA gauge in the 
SU(2) lattice QCD with $16^3 \times 4$.(a) confinement phase 
($\beta =2.2$), (b) deconfinement phase ($\beta =2.4$).
There appears the global network of monopole currents 
in the confinement phase.
}
                    \label{fig:radish2}
     \end{figure}

\section{Analytical Proof of Abelian Dominance for Confinement}

As long as confinement, monopole dominance seems trivial 
if abelian dominance holds, 
because the electric current $j_\mu$ does not contribute to 
the electric confinement. 
In the confinement physics, 
the nontrivial phenomenon observed in the MA gauge 
is abelian dominance. 
In this section, we analytically prove 
abelian dominance on the string tension in the MA gauge, 
considering the off-diagonal gluon properties.$^{2,6}$
In the lattice formalism, the SU(2) link variable is factorized 
as $U_\mu(s)=M_\mu(s)u_\mu(s)$, according to the Cartan decomposition 
$G \simeq G/H \times H$. 
Here, $u_\mu(s)\equiv \exp\{i\tau^3\theta^3_\mu(s)\}\in H$ 
denotes the abelian link variable, and the off-diagonal factor 
$M_\mu(s)\in G/H$ is parameterized as 
\be
M_\mu(s)\equiv e^{i\{\tau^1\theta^1_\mu(s)+\tau^2\theta^2_\mu(s)\}}
=\left( {\matrix{
{\rm cos}{\theta_\mu}(s) & -{\rm sin}{\theta_\mu}(s) e^{-i\chi_\mu(s)} \cr
{\rm sin}{\theta_\mu}(s) e^{i\chi_\mu(s)} & {\rm cos}{\theta_\mu}(s)
}} \right). 
\ee

In the MA gauge, the {\rm diagonal element} $\cos \theta_\mu(s)$ 
in $M_\mu(s)$ is maximized by the gauge transformation 
as large as possible; for instance, the abelian projection rate 
is almost unity as 
$R_{\rm Abel}=\langle\cos \theta_\mu(s)\rangle_{\rm MA}\simeq 0.93$ 
at $\beta=2.4$. 
Then, the MA gauge fixing provides the two remarkable properties on 
the off-diagonal element $e^{i\chi_\mu(s)}\sin\theta_\mu(s)$ 
in $M_\mu(s)$. 

\noindent
\begin{minipage}{.5cm}
(1)~\\
\end{minipage}
\begin{minipage}{11.3cm}
the off-diagonal amplitude $|\sin\theta_\mu(s)|$ 
is forced to be small in the MA gauge, 
which allows the approximate treatment on the off-diagonal element.
\end{minipage}
\noindent
\begin{minipage}{.5cm}
(2)~\\~\\
\end{minipage}
\begin{minipage}{11.3cm}
the off-diagonal angle variable $\chi_\mu(s)$ is not 
constrained by the MA gauge-fixing condition at all, 
and tends to be a random variable.
\end{minipage}

\noindent
Hence, $\chi_\mu(s)$ can be regarded as a {\it random angle variable} 
on the treatment of $M_\mu(s)$ in the MA gauge 
in a good approximation.$^{2,6}$ 

In calculating the Wilson loop 
$\langle W_C[U_\mu(\cdot)]\rangle =
\langle{\rm tr}\Pi_C\{M_\mu(s)u_\mu(s)\}\rangle$, 
we take the random-variable approximation for the 
off-diagonal angle variable $\chi_\mu(s)$ in the MA gauge, and then 
the integral of $e^{i\chi_\mu(s)}$ on $\chi_\mu(s)$ vanishes as 
\be
\langle e^{i\chi_\mu(s)}\rangle 
\simeq \int_0^{2\pi} d\chi_\mu(s)\exp\{i\chi_\mu(s)\}=0.
\ee
Thus, the {\it off-diagonal factor} 
$M_\mu(s)$ appearing in 
$\langle W_C[U_\mu(\cdot)]\rangle$ is simply reduced as a $c$-number factor, 
$
M_\mu(s) \rightarrow \cos \theta_\mu(s) \ {\bf 1},
$
and therefore the SU(2) link variable $U_\mu(s)$ in 
$\langle W_C[U_\mu(\cdot)]\rangle$ 
is reduced to a {\it diagonal matrix,}
\be
U_\mu(s)\equiv M_\mu(s)u_\mu(s)
\rightarrow 
\cos \theta_\mu(s) u_\mu(s).
\ee
Then, for the $I \times J$ rectangular $C$, the Wilson loop 
$W_C[U_\mu(\cdot)]$ in the MA gauge is approximated as 
\bea
\langle W_C[U_\mu(\cdot)]\rangle 
	&\equiv&
	\langle{\rm tr}\Pi_{i=1}^L U_{\mu_i}(s_i)\rangle
	\simeq 
	\langle\Pi_{i=1}^L \cos \theta_{\mu_i}(s_i) \cdot 
	{\rm tr} \Pi_{j=1}^L u_{\mu_j}(s_j)\rangle_{\rm MA} \nonumber\\
	&\simeq&
	\langle\exp\{\Sigma_{i=1}^L 
		\ln (\cos \theta_{\mu_i}(s_i))\}\rangle_{\rm MA} 
	\ \langle W_C[u_\mu(\cdot)]\rangle_{\rm MA},
\eea
where $L\equiv 2(I+J)$ denotes the perimeter length and 
$W_C[u_\mu(\cdot)]\equiv {\rm tr}\Pi_{i=1}^L u_{\mu_i}(s_i)$ 
the abelian Wilson loop.
Replacing $\sum_{i=1}^L \ln \{\cos(\theta_{\mu_i}(s_i))\}$ 
by its average \\
$L \langle \ln \{\cos \theta_\mu(\cdot)\} \rangle_{\rm MA}$
in a statistical sense, we derive a simple estimation as$^{2,6}$ 
\bea
W_C^{\rm off}\equiv 
\langle W_C[U_\mu(\cdot)]\rangle/\langle W_C[u_\mu(\cdot)]\rangle_{\rm MA}
\simeq \exp\{L\langle \ln(\cos \theta_\mu(\cdot))\rangle_{\rm MA}\}
					\label{eqn:9}
\eea
for the {\it contribution of the off-diagonal 
gluon element to the Wilson loop}.
From this analysis, the contribution of off-diagonal gluons 
to the Wilson loop is expected to obey the {\it perimeter law} 
in the MA gauge for large loops, where the statistical 
treatment would be accurate.

We show 
the off-diagonal contribution 
$W_C^{\rm off}\equiv \langle W[U_\mu(\cdot)]\rangle/
\langle W[u_\mu(\cdot)]\rangle_{\rm MA}$ 
to the Wilson loop in the lattice QCD simulation with $\beta=2.4$ in Fig.2. 
In the MA gauge,  $W_C^{\rm off}$ seems to obey the 
{\it perimeter law} for the large loop, 
and is well reproduced by the estimation of Eq.(\ref{eqn:9})
with the {\it microscopic input} 
as $\langle \ln \{\cos\theta_\mu(s)\} \rangle_{\rm MA}\simeq -0.082$ 
for $\beta=2.4$.
From Eq.(\ref{eqn:9}), 
the off-diagonal contribution to the string tension vanishes as 
\bea
\sigma_{\rm SU(2)}-\sigma_{\rm Abel} 
\simeq 
-2 \langle \ln \{\cos\theta_\mu(s)\} \rangle_{\rm MA}
\lim_{R,T \rightarrow \infty} {R+T \over RT}=0.
\eea
Thus, {\it abelian dominance for the string tension}, 
$\sigma_{\rm SU(2)}=\sigma_{\rm Abel}$, 
can be proved in the MA gauge within 
the random-variable approximation for 
the off-diagonal angle variable $\chi_\mu(s)$, 
although the {\it finite size effect} on $R$ and $T$ 
in the Wilson loop leads to the deviation between 
$\sigma_{\rm SU(2)}$ and $\sigma_{\rm Abel}$  as
$\sigma_{\rm SU(2)} > \sigma_{\rm Abel}$.$^{2,6}$
\begin{figure}[htb]
\hspace{3cm}\epsfig{figure=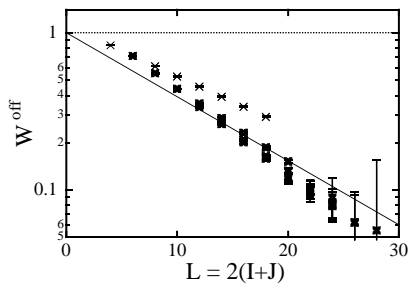,height=4cm}
\caption{
The comparison between the lattice data and 
the analytical estimation of $W_C^{\rm off}$ 
$\equiv$ 
$\langle W_C[U_\mu(\cdot)] \rangle/
\langle W_C[u_\mu(\cdot)]\rangle$ 
as the function of the perimeter $L \equiv 2(I+J)$ in the MA gauge. 
The cross ($\times$) denotes the lattice date 
at $\beta=2.4$, 
and the straight line denotes the theoretical estimation of 
$W_C^{\rm off} = 
\exp\{L \langle \ln(\cos \theta_\mu(s))\rangle_{\rm MA} \}$ 
with the microscopic input 
$\langle \ln \{\cos \theta_\mu(s)\}\rangle_{\rm MA} \simeq -0.082$ 
at $\beta=2.4$.
The off-diagonal gluon contribution $W_C^{\rm off}$
seems to obey the {\it perimeter law} for $I, J \ge 2$. 
}
                    \label{fig:off-w}
\end{figure}

\section{Origin of Abelian Dominance :
   Effective Charged-Gluon Mass induced in MA Gauge}

In this section, we study the origin of abelian dominance 
for NP-QCD in the MA gauge in terms of the {\it generation of the 
effective mass $m_{ch}$ of the off-diagonal (charged) gluon 
by the MA gauge fixing} in the QCD generating functional 
as$^{2,12}$ 
\bea
Z_{\rm QCD}^{\rm MA} &=& \int DA_\mu \exp\{iS_{\rm QCD}[A_\mu ]\} 
\delta (\Phi _{\rm MA}^\pm [A_\mu ])\Delta _{\rm PF}[A_\mu ] \cr
&\simeq& 
\int DA_\mu ^3 \exp\{iS_{\rm eff}[A_\mu ^3]\}
\int DA_\mu ^\pm \exp\{i\int d^4x \ 
m_{ch}^2 A_\mu ^+A^\mu _- \} {\cal F}[A_\mu ],
\eea
where $\Delta _{\rm FP}$ is the Faddeev-Popov determinant, 
$S_{\rm eff}[A_\mu ^3]$ the abelian effective action 
and ${\cal F}[A_\mu ]$ a smooth functional. 
If the MA gauge fixing induces the effective mass 
$m_{ch}$ of off-diagonal (charged) gluons, 
the charged gluon propagation is limited within 
the short-range region as $r \mathop{^<}_{\!\!\!\!\!\sim} m_{ch}^{-1}$, 
and hence off-diagonal gluons cannot contribute 
to the long-distance physics in the MA gauge, which 
provides the origin of abelian dominance for NP-QCD.  

Here, using the SU(2) lattice QCD in the Euclidean metric, 
we study the gluon propagator 
$G_{\mu \nu }^{ab} (x-y) \equiv \langle A_\mu ^a(x)A_\nu ^b(y)\rangle$ 
in the MA gauge.$^{2,13}$ 
As for the residual U(1)$_3$ gauge symmetry, 
we impose the U(1)$_3$ Landau gauge fixing 
to extract most continuous gauge configuration and 
to compare with the continuum theory.
The gluon field $A_\mu^a(x)$ is extracted 
from the link variable as 
$U_\mu(s)={\rm exp}(iaeA_\mu^a(s) \frac{\tau^a}{2})$.
Here, the scalar combination 
$G_{\mu\mu}^a(r)\equiv \sum^4_{\mu=1}\langle 
A_\mu^{~a}(r)A_\mu^{~a}(0)\rangle~(a=1,2,3)$ 
is useful to observe the interaction range of the gluon,
because it depends only on the four-dimensional Euclidean 
radial coordinate $r \equiv (x_\mu x_\mu)^{{1 \over 2}}~$.

We calculate the gluon propagator $G_{\mu \mu }^a(r)$ in the MA gauge 
using the SU(2) lattice QCD 
with $12^3 \times 24$ and $2.2 \le \beta \le 2.4$.
In the MA gauge, 
the off-diagonal (charged) gluon propagates only within the 
short-range region $r \mathop{^<}_{\!\!\!\!\!\sim} 0.4$ fm, 
so that it cannot contribute to the long-range physics. 
On the other hand, the diagonal gluon propagates over the long distance 
and influences the long-range physics.
Thus, we find {\it abelian dominance for the gluon propagator 
in the MA gauge}, and this is the {\it origin of abelian dominance 
for the long-distance physics or NP-QCD.}$^{2,13}$

Since the propagator of the massive vector boson with mass $M$ 
asymptotically behaves as 
the Yukawa-type function $G_{\mu\mu}(r)\simeq 
{3 \sqrt{M} \over (2\pi)^{3/2}}\cdot{\exp(-Mr) \over r^{3/2}}$,
the effective mass $m_{ch}$ of the charged gluon 
can be evaluated from the slope of the logarithmic plot of 
$r^{3/2}G_{\mu\mu}^{+-}(r)\sim \exp(-m_{ch}r)$ as shown in 
Fig.3. 
The charged gluon behaves as a massive particle 
at the long distance, $r \mathop{^>}_{\!\!\!\!\!\sim} 0.4$ fm.
We obtain the {\it effective mass of the charged gluon} 
as $m_{ch} \simeq 0.94~{\rm GeV}$,$^{2,13}$ 
which provides the {\it critical scale on abelian dominance.}

\ 

\begin{figure}[htb]
\epsfig{figure=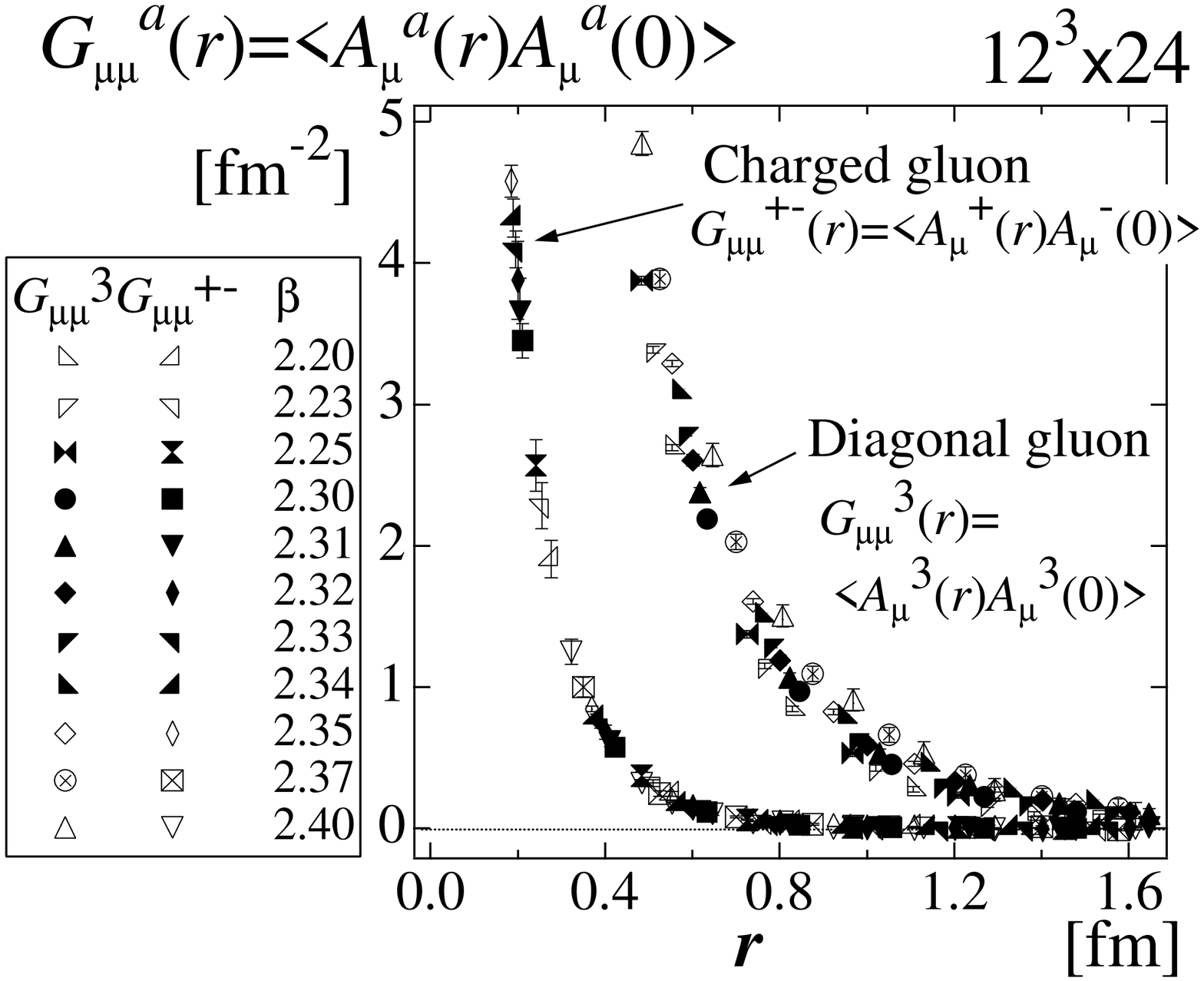,height=5.3cm}
\epsfig{figure=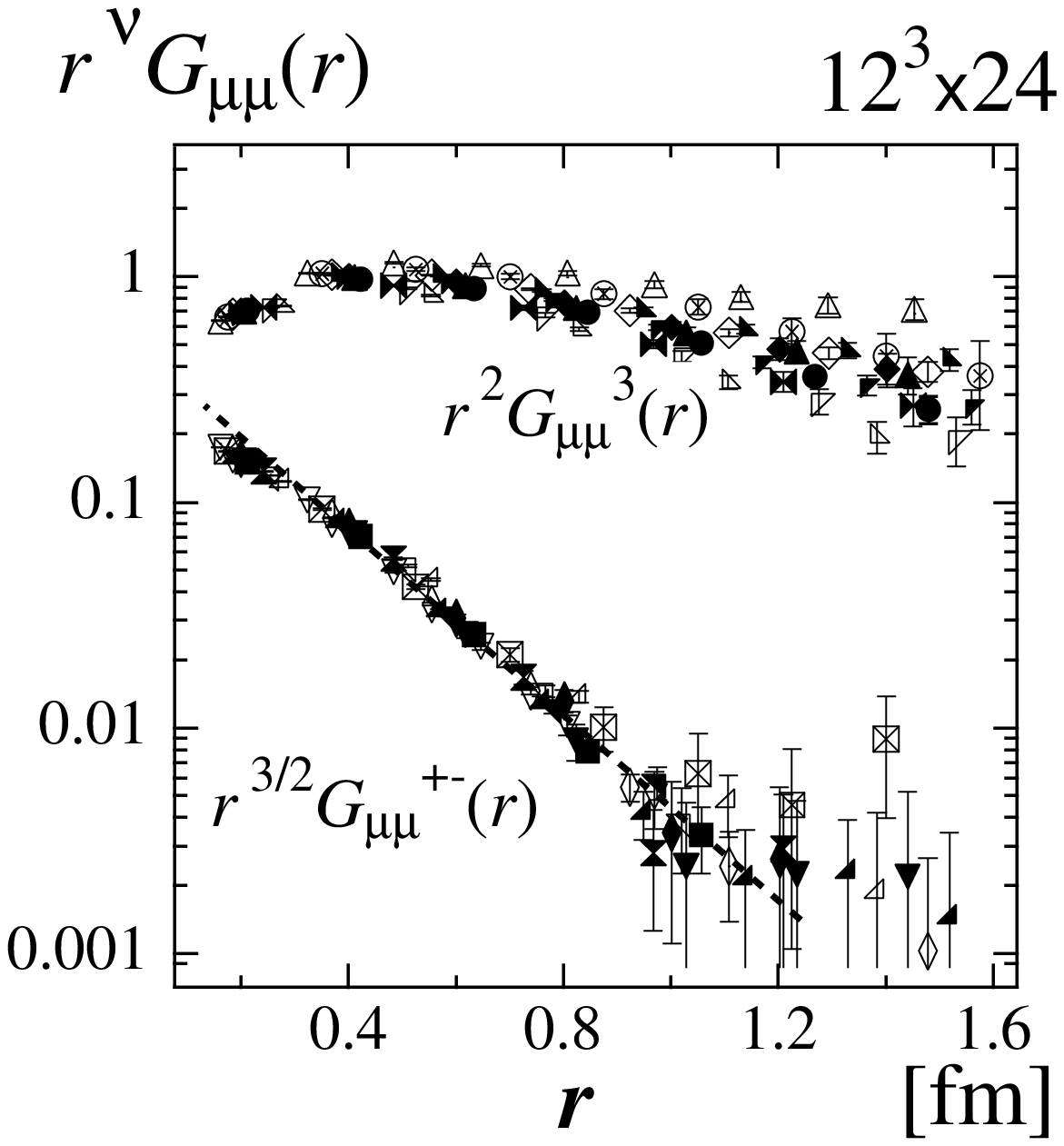,height=5.3cm}\\
$~\hspace{3.2cm}$(a)\hspace{5.4cm}(b)
          \caption{ (a) The scalar correlation $G_{\mu \mu }^a(r)$ 
of the gluon propagator as the function of the 4-dimensional 
distance $r$ in the MA gauge in the SU(2) lattice QCD with 
$12^3 \times 24$ and $2.2 \le \beta \le 2.4$.
In the MA gauge, 
the off-diagonal (charged) gluon propagates within the 
short-range region $r \mathop{^<}_{\!\!\!\!\!\sim} 0.4$ fm, 
and cannot contribute to the long-range physics. 
(b) The logarithmic plot for 
the scalar correlation $r^\nu G_{\mu \mu }^a(r)$. 
The charged-gluon propagator behaves as 
the Yukawa-type function, 
$G_{\mu \mu } \sim {\exp(-m_{ch}r) \over r^{3/2}}$. 
The effective mass  of the charged gluon 
can be estimated as $m_{ch}\simeq 0.94{\rm GeV}$ 
from the slope of the dotted line. }
                    \label{fig:radish2}
\end{figure}

\section{Dual Wilson Loop, 
Inter-Monopole Potential and Evidence of
Dual Higgs Mechanism (Monopole Condensation)}

In this section, we study the dual Higgs mechanism by 
monopole condensation in the NP-QCD vacuum 
in the field-theoretical manner. 
Since QCD is described by the ``electric variable'' as 
quarks and gluons, 
the ``electric sector'' of QCD has been well studied with 
the Wilson loop or the inter-quark potential, however, 
the ``magnetic sector'' of QCD is hidden and still unclear. 
To investigate the magnetic sector directly,
it is useful to introduce the ``dual (magnetic) variable'' 
as the {\it dual gluon} $B_\mu $, 
similarly in the dual Ginzburg-Landau (DGL) theory.$^{5,14-17}$
The dual gluon $B_\mu $ is the dual partner of the diagonal gluon 
and directly couples with the magnetic current $k_\mu $.

Here, we concentrate ourselves to the monopole part 
in the MA gauge, which holds the essence of NP-QCD. 
Since the monopole part does not include the electric current as 
$\partial_\mu F^{\mu\nu}=j^\nu \simeq 0$, 
the dual gluon $B_\mu $ can be introduced 
as the regular field satisfying 
$\partial_\mu B_\nu - \partial_\nu B_\mu={^*\!F}_{\mu\nu}$ 
and the dual Bianchi identity, 
$
{\partial^{\mu}} {^*\!(}\partial \land B)_{\mu\nu}=0.
$
In terms of the dual Higgs mechanism, 
the inter-monopole potential is expected to be short-range 
Yukawa-type, and the dual gluon $B_\mu $ becomes massive 
in the monopole-condensed vacuum.
To examine the inter-monopole potential, 
we define {\it the dual Wilson loop} $W_D$ as the line-integral of 
the dual gluon $B_\mu$ along a loop $C$,$^{2,12,18}$
\be
W_D(C) \equiv \exp\{i{e \over 2}\oint_C dx_\mu B^\mu \}=
\exp\{i{e \over 2}\int\!\!\!\int d\sigma_{\mu\nu}{^*\!F}^{\mu\nu}\},
\ee
which is the {\it dual version of the abelian Wilson loop}
$W_{\rm Abel}(C) \equiv $
~\\
$\exp\{i{e \over 2}\oint_C dx_\mu A^\mu \}=
\exp\{i{e \over 2}\int\!\!\!\int d\sigma_{\mu\nu}{F}^{\mu\nu}\}$.
Here, we have set the test monopole charge as $e/2$
considering the duality correspondence.
The potential between the monopole and the anti-monopole 
is derived from the dual Wilson loop as 
\be
V_{M}(R) = -\lim_{T \rightarrow  \infty} {1 \over T}\ln 
\langle W_D(R,T) \rangle.
\ee
Using the SU(2) lattice QCD in the MA gauge,
we study the dual Wilson loop and 
the inter-monopole potential in the monopole part.
The dual Wilson loop $\langle W_D(R,T) \rangle$ seems to 
obey the {\it perimeter law} as 
$\langle W_D(R,T) \rangle \sim \exp\{-c(R+T)\}$ for large $R$, $T$. 
As shown in Fig.4, the inter-monopole potential 
is short-ranged and flat in comparison with 
the inter-quark potential.

\ 

\begin{figure}[htb]
\hspace{3cm}
\epsfig{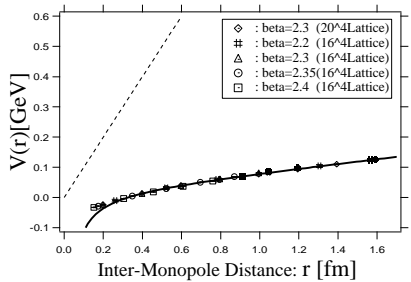}

          \caption{
The inter-monopole potential $V_M(r)$ 
extracted from the dual Wilson loop 
$\langle W_D(I,J) \rangle$ in the monopole part 
in the MA gauge.
Here, $r$ denotes the 3-dimensional distance 
between the monopole and the anti-monopole. 
For comparison, we add the linear inter-quark potential 
denoted by the dashed line. 
The solid curve denote the Yukawa potential adding 
the finite-size effect of the dual Wilson loop. 
}

     \end{figure}

Except for the short distance, the inter-monopole potential 
can be fitted by the Yukawa potential 
$V_M(r) = -{{(e/2)}^2 \over 4\pi}{e^{-m_Br} \over r}+{cr \over Ta}$, 
where the second term is the correction appearing as 
the finite-size effect of the dual Wilson loop.
The dual gluon mass is estimated as 
$m_B \simeq {\rm 0.5GeV}$, 
which is consistent with the DGL theory.$^{5,14-17}$
{\it The mass generation of the dual gluon $B_\mu $ 
can be regarded as the direct evidence of the dual Higgs mechanism 
by monopole condensation at the infrared scale 
in the NP-QCD vacuum.}$^{2,12,18}$

Thus, the lattice QCD in the MA gauge exhibits abelian dominance
and monopole condensation in the infrared region, and leads 
to the dual Ginzburg-Landau theory as the infrared effective
theory directly based on QCD$^{2,12,19}$ (See Fig.5).

\section*{Acknowledgments}
H.S. and H.I. acknowledge Dr. K.Tsushima for his warm hospitality at CSSM. 

\begin{figure}[htb]
\epsfig{figure=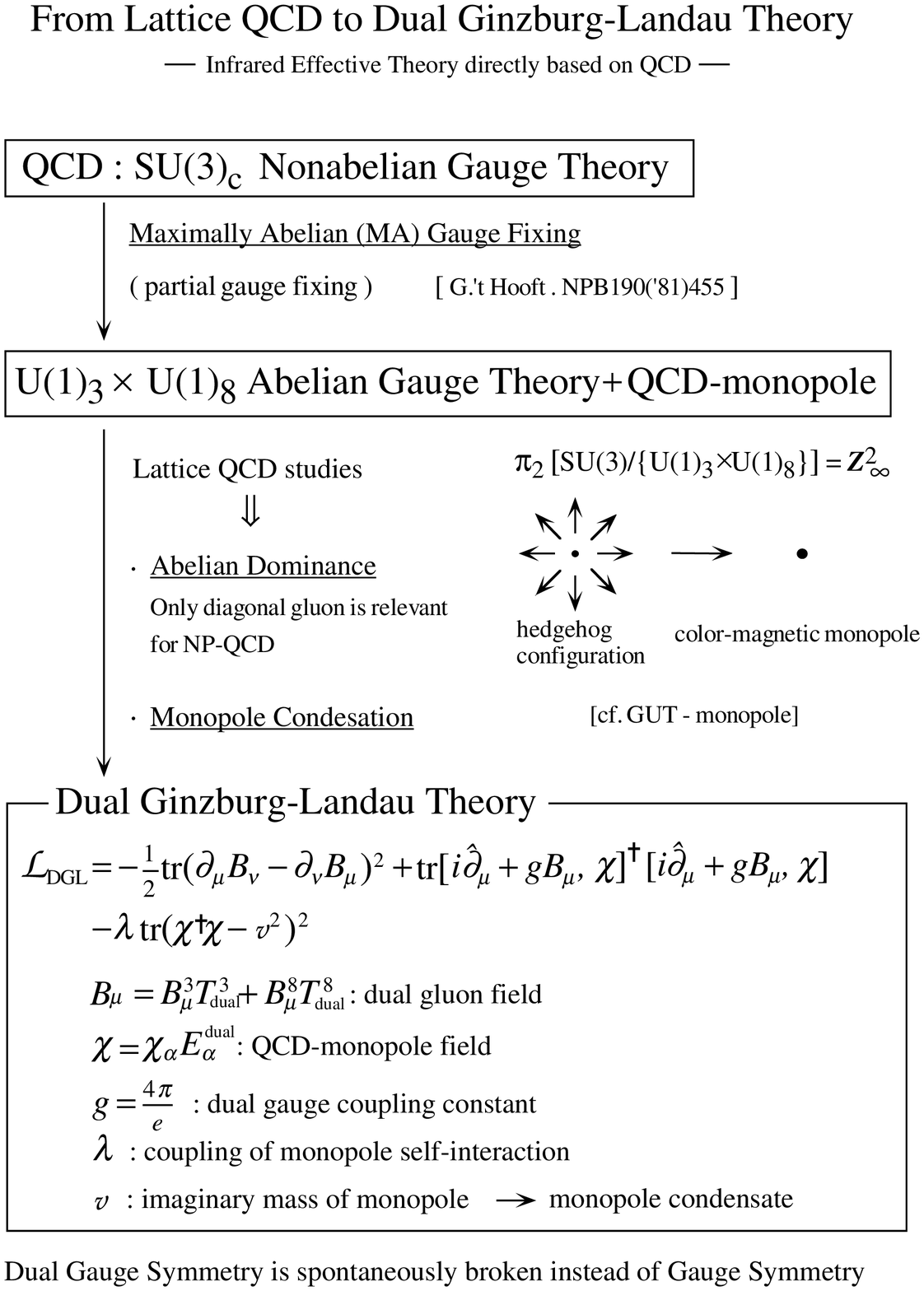,height=15.5cm}
\caption{
Construction of the dual Ginzburg-Landau (DGL) theory 
from the lattice QCD in the maximally abelian (MA) gauge. }
\end{figure}

\end{document}